\documentclass[prl, twocolumn,amsmath,superscriptaddress]{revtex4}
\usepackage{graphicx}% Include figure files
\usepackage{dcolumn}% Align table columns on decimal point
\usepackage{bm}% bold math
\usepackage{mathbbol}

\newcommand{\mathcommand}[3][0]{\newcommand{#2}[#1]{\ensuremath{#3}}}

\newcommand{\ts}[2]{{#1}_{\textnormal{#2}}} %A math symbol with a text subscript
\newcommand{\tsc}[2]{{#1}_{\textsc{#2}}} %A math symbol with a text subscript, small caps

\newcommand{\abbrev}[1]{{\scshape\lowercase{#1}}}  % see LaTeX Companion, p. 169

\renewcommand{\vec}[1]{\ensuremath{\mathbf{#1}}}

\newcommand{\be}{\begin{equation}}
\newcommand{\ee}{\end{equation}}
\newcommand{\splitstart}{\begin{equation}\begin{split}}
\newcommand{\splitstop}{\end{split}\end{equation}} %Why doesn't this work?!?

\newcommand{\refeq}[1]{Eq.~\eqref{#1}}

\newcommand{\reffig}[1]{Fig.~\ref{#1}}

\mathcommand[1]{\smallket}{|#1\rangle}
\mathcommand[1]{\smallbra}{\langle #1|}
\mathcommand[1]{\bigket}{\bigl|#1\bigr\rangle}
\mathcommand[1]{\bigbra}{\bigl\langle #1\bigr|}
\mathcommand[1]{\biggket}{\biggl|#1\biggr\rangle}
\mathcommand[1]{\biggbra}{\biggl\langle #1\biggr|}
\mathcommand[1]{\ket}{\left| #1 \right\rangle}  %a Ket
\mathcommand[1]{\lket}{\bigl| #1 \bigr)}  %a Liouville Ket
\mathcommand[1]{\bra}{\left\langle #1 \right|}  %a Bra
\mathcommand[1]{\lbra}{\bigl( #1 \bigr|}  %a Bra
\newcommand{\latinfont}[1]{#1}  % I think this is what the Chi. Man. of Style suggests

\newcommand{\etal}{\latinfont{et al.}} %Note period is omitted to be used in BibTeX

\mathcommand{\te}{\text{e}}
\mathcommand{\vactext}{\text{vac}}
\mathcommand{\vacket}{\ket\vactext}

\newcommand{\QD}{\abbrev{QD}}
\newcommand{\QDs}{\abbrev{QD}s}
\newcommand{\FID}{\abbrev{FID}}

\newcommand{\pdf}{\abbrev{pdf}}
\mathcommand{\am}{\langle m(\vec{r})\rangle}

\begin{document}
\title{Pulsed Nuclear Pumping and Spin Diffusion in a Single Charged Quantum Dot}

\newcommand\ginzton{E. L. Ginzton Laboratory,
        Stanford University,
        Stanford, California 94305, USA}
\newcommand\NII{National Institute of Informatics,
        Hitotsubashi 2-1-2, Chiyoda-ku,
        Tokyo 101-8403, Japan}
\newcommand\wurzburg{Technische Physik, Physikalisches Institut,
        Wilhelm Conrad R\"{o}ntgen Research Center for Complex Material Systems,
        Universit\"{a}t W\"{u}rzburg,
        Am Hubland, D-97074 W\"{u}rzburg, Germany}

\author{Thaddeus~D.~Ladd}
    \altaffiliation{Currently at HRL Laboratories, LLC,
    3011 Malibu Canyon Rd., Malibu, CA 90265.
     Electronic address:
    \texttt{tdladd@gmail.com}}
    \affiliation{\ginzton}\affiliation{\NII}
\author{David~Press}
    \affiliation\ginzton
\author{Kristiaan~De~Greve}
    \affiliation\ginzton
\author{Peter~L.~McMahon}
    \affiliation\ginzton
\author{Benedikt~Friess}
    \affiliation\ginzton\affiliation\wurzburg
\author{Christian~Schneider}
    \affiliation\wurzburg
\author{Martin~Kamp}
    \affiliation\wurzburg
\author{Sven~H\"{o}fling}
    \affiliation\wurzburg
\author{Alfred~Forchel}
    \affiliation\wurzburg
\author{Yoshihisa~Yamamoto}
    \affiliation{\ginzton}\affiliation{\NII}

\begin{abstract}
We report the observation of a feedback process between the
nuclear spins in a single charged quantum dot and its trion
transition, driven by a periodic sequence of optical pulses.
The pulse sequence intersperses off-resonant ultrafast pulses
for coherent electron-spin rotation and resonant narrow-band
optical pumping.  The feedback manifests as a hysteretic
triangle-like pattern in the free-induction-decay of the single
spin. We present a simple, quasi-analytic numerical model to
describe this observation, indicating that the feedback process
results from the countering effects of optical nuclear pumping
and nuclear spin-diffusion inside the quantum dot. This effect
allows dynamic tuning of the electron Larmor frequency to a
value determined by the pulse timing, potentially allowing more
complex coherent control operations.
\end{abstract}
\maketitle

Optically controlled quantum dots (\QDs) are in many ways
similar to atomic systems, and are therefore often regarded as
strong candidates for solid-state quantum information
processing. However, one key feature distinguishing \QDs\ in
group \abbrev{III-V} semiconductors from atomic systems is the
presence of a large nuclear-spin ensemble~\cite{sp08}.
%Coherent processes in quantum dots show strong promise for
%several applications related to quantum information, since
%quantum dots show atom-like behavior without the need for
%cooling and trapping.  However, the effects of nuclear spin
%ensembles in quantum dots based on III-V semiconductors provide
%a large difference between these atom-like complexes and actual
%atoms~\cite{sp08}.
Nuclear spins cause adverse effects such as inhomogeneous
broadening and non-Markovian decoherence processes. However,
nuclear spins may play useful roles as well.  Although methods
to use \QD\ nuclear spins directly as a quantum memory remain
challenging due to the difficulty of achieving sufficiently
high levels of nuclear polarization, nuclear spins may provide
novel methods for the dynamic \emph{tuning} and \emph{locking}
of electron spin resonances for electrons trapped in \QDs.

Several examples of manipulating nuclei to improve electron
spin coherence have recently been observed.  In electrically
controlled double \QDs, transition processes between electron
singlet and triplet states allow the manipulation of interdot
nuclear spin polarizations, improving coherent
control~\cite{vandersypen_singlet-triplet,Reilly_Zamboni_Science,new_yacoby}.
In single \QDs\ under microwave control, nuclear effects
dynamically tune the electron spin resonance to the applied
microwave frequency~\cite{delftlock}.  Tuning effects are also
observed in two-color continuous-wave (CW) laser experiments,
in which the appearance of coherent electronic effects such as
population trapping are modified by nonlinear feedback
processes with nuclear spins~\cite{sham_cpt,imamoglu_cpt}.
Finally, nuclear spins have been shown to dynamically bring
ensembles of inhomogeneous \QDs\ into spin-resonance with a
train of ultrafast
pulses~\cite{greilich_nuclear,Reinecke_nuclear}.

Here, we describe a related but different manifestation of the
non-Markovian dynamics occurring between a single electron in a
\QD\ and the nuclear bath with which it interacts, with new
possibilities for use in controlling nuclear effects.  The
effect occurs when measuring the familiar ``free-induction
decay" (\FID) of a single spin in a single \QD\ under pulsed
control. The Larmor frequency of the electron spin is
dynamically altered by the hyperfine interaction with \QD\
nuclei; the nuclear polarization is in turn altered by the
measurement results of the \FID\ experiment. The result is a
feedback loop in which the nuclear hyperfine field stabilizes
to a value determined by the timing of the pulse sequence.  In
what follows, we show the experimental manifestation of this
feedback loop and present a numerical model for the effect.

The \FID\ or Ramsey interferometer experiment proceeds by
tipping a single electron spin from the pole of its Bloch
sphere onto the equator with a $\pi/2$ rotation, allowing it to
precess for a time $\tau$, and then tipping it with a second
$\pi/2$ rotation into a state with polarization depending
sinusoidally on $\tau$.  We accomplish this experiment in a
single, charged, self-assembled InAs \QD\ using periodically
applied pairs of ultrafast optical pulses, interspersed with
optical pumping. The molecular-beam-epitaxially grown \QDs\ are
centered in a planar microcavity with 24 and 5 pairs of
GaAs/AlAs $\lambda$/4 mirror pairs below and above the cavity,
respectively. The cavity quality factor was roughly $Q\approx
200$, and the single-sided design of the cavity directed most
emission towards the collection optics.  A layer of Si dopants
placed 10~nm below the \QDs\ served to probablistically dope
the \QDs\ with electrons. Several prior experiments have shown
that single ultrafast pulses achieve coherent, single-spin
rotations in charged
\QDs~\cite{awschsingle_etal,pressnature,kimarxiv}. The rotation
occurs on picosecond timescales, which is effectively
instantaneous in comparison to electron-spin decoherence times
or nuclear spin dynamics. The optical pumping in our
experiment, previously described in
Ref.~\onlinecite{pressnature},  is accomplished via a trion
transition.  It serves to repeatedly re-initialize the spin
into one particular polarization state prior to the ultrafast
pulses.  Spin measurement is accomplished by
single-photon-counting during the optical pumping step; the
photon count-rate is proportional to the probability of
exciting a trion state.  This count-rate averaged over many
sequential experiments is expected to vary sinusoidally with
$\tau$, resulting in ``Ramsey fringes," and to decay on a
timescale called $T_2^*$ due to dephasing or decoherence
effects.

%There are two differences between the present experimental
%set-up and previous work~\cite{pressnature}.  First, the sample
%is different: in the present study a metal mask isolates a
%single quantum dot rather than an etched mesa, and the quantum
%dot's emission is directed by a distributed-Bragg-reflector
%microcavity ($Q\sim 200$). Second, and more relevantly,
In contrast to earlier work~\cite{pressnature}, the resonant
laser used to optically initialize the single spin is gated off
during the free precession of the electron spin using an
electro-optic modulator.  This eliminates the previously
observed source of decay of Ramsey fringes, leading to an
expected longer $T_2^*$ decay. In particular, a Markovian
picture of nuclear spin noise processes predicts a Gaussian
decay with $T_2^*$ timescale on the order of a few nanoseconds.

\begin{figure}[t]
\includegraphics[width=\columnwidth]{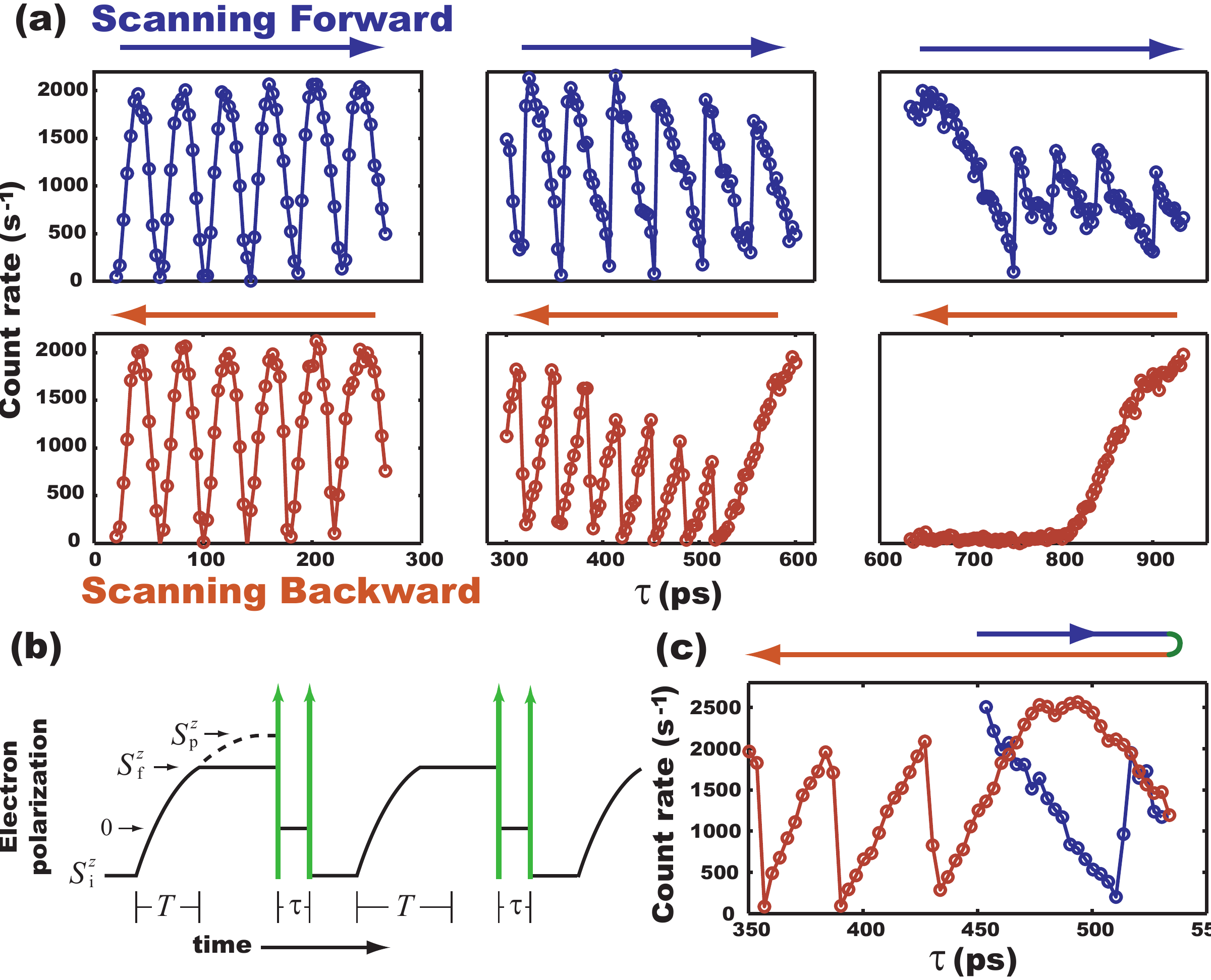}
\caption{(Color online) (a) Experimental Ramsey fringe count-rate
as a function of two-pulse time delay $\tau$,
measured using the techniques of Ref.~\onlinecite{pressnature}.
This data was taken at a magnetic field of 4~T, but similar effects
are seen at different magnetic fields.
(b)
Average electron polarization as a result of the periodic pulse sequence used to generate this data.
Optical pumping increases the polarization for a duration $T=26$~ns.
The saturation polarization, which would be reached for $T\rightarrow\infty$ (dashed line)
is $\ts{S}{p}^z$; in time $T$ only polarization $\ts{S}{f}^z$ is reached.
After pumping and a short delay, a picosecond pulse indicated by a green (grey) arrow
nearly instantaneously rotates
the electron spin to the equator of the Bloch sphere ($\langle S^z \rangle=0$); a time
$\tau$ later a second pulse rotates the spin to achieve electron polarization $\ts{S}{i}^z$,
depending on the amount of Larmor
precession between the pulses.   The theoretical count-rate $C(\omega,\tau)$ of \refeq{countrate}
is found as $\ts{S}{f}^z-\ts{S}{i}^z$ in steady-state conditions.  (c) Experimental Ramsey
fringe count-rate as $\tau$ is continuously scanned forward and then backward, showing clear hysteresis.
}
\label{data}
\end{figure}

However, such a Gaussian decay was not observed.
Figure~\ref{data} shows the result of the \FID\ experiment. The
top three traces show the fringes seen as $\tau$ is scanned in
the forward direction (increasing the delay $\tau$ by moving a
retroreflector outwards with a scanning stage), and the bottom
three correspond to scanning in the backward direction
(reducing the delay $\tau$ by moving the retroreflector
inwards).  The oscillatory fringes, rather than decaying,
evolve into a sawtooth pattern at high values of
$\tau$~\footnote{Similar data is seen in similar experiments
performed elsewhere; see the top trace of Fig. 3b of
Ref.~\onlinecite{kimarxiv}.}, and show hysteresis depending on
the direction in which $\tau$ is changed.  The break between
traces in Fig.~\ref{data}(a) is due to the need to manually
move the scanning stage.

This data is the result of two competing processes: changes in
the average nuclear hyperfine shift $\omega$ due to trion
emission, in conjunction with the motion of that magnetization
due to spin diffusion. In what follows, we first qualitatively
describe these physical processes and explain how they lead to
our data, and then we present equations to formally model the
dynamics quantitatively.

One important assumption is a separation of dynamics into three
very distinct timescales.  The fastest timescale is the pulse
sequence and resulting electron-spin dynamics, repeated
continuously with a repetition period of 143~ns, shown in
\reffig{data}(b).  This is much faster than the nuclear
dynamics we consider, which are presumed to occur on
millisecond timescales.  Finally, the averaging timescale of
the measurement is much longer still, on the order of several
seconds, allowing the nuclei substantial time to reach
quasi-equilibrium.

\begin{figure}
\includegraphics[width=0.7\columnwidth]{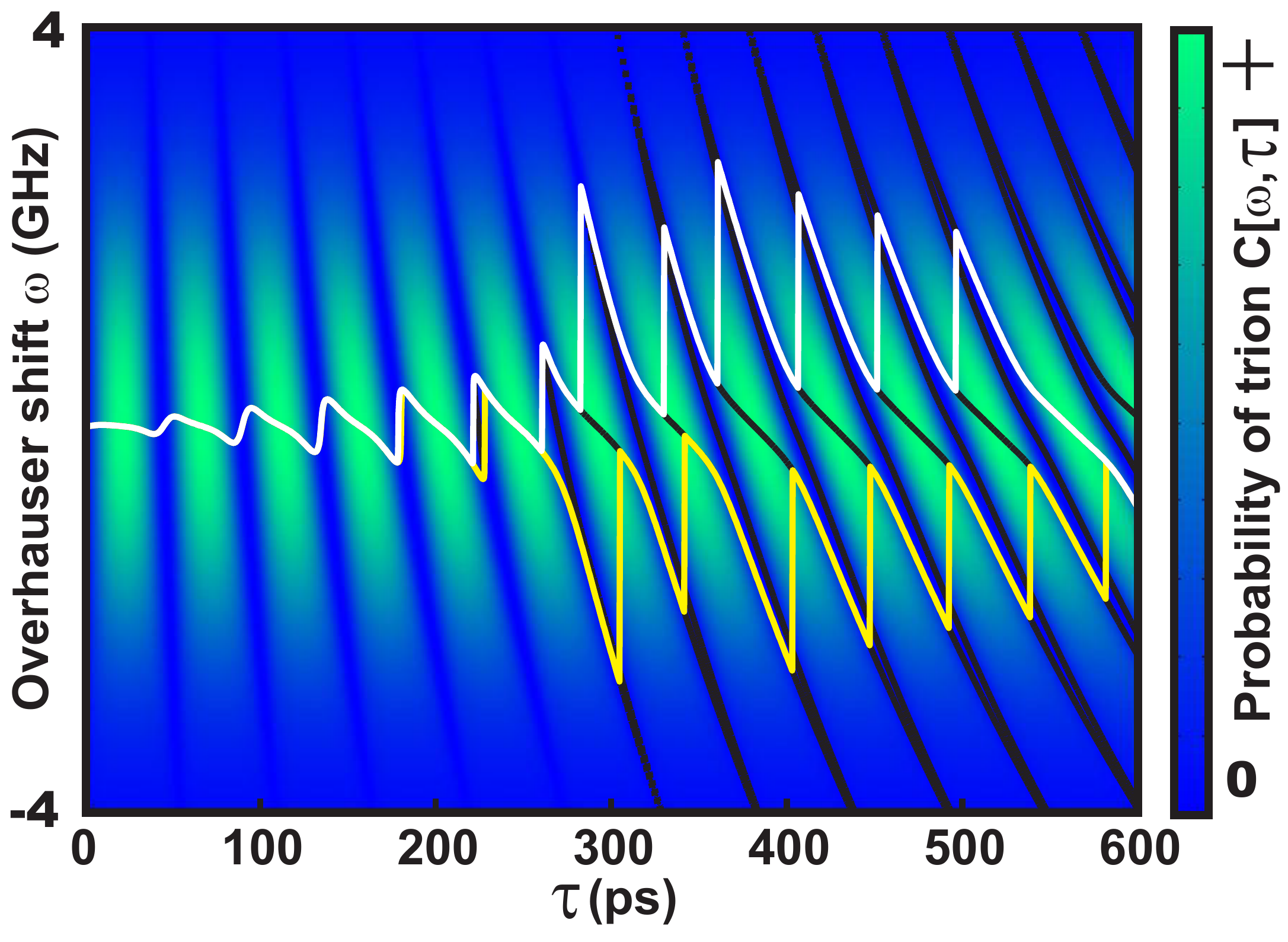}
\caption{(Color online) Count-rate $C(\omega,\tau)$ as a function of
Overhauser shift $\omega$ and two-pulse delay $\tau$.  The lighter grey
or greener areas indicate where a higher count-rate is expected.
Oscillations in the horizontal directions at frequency $\omega_0+\omega$
are due to Ramsey interference; the Gaussian envelope in the vertical
direction is due to the reduction of optical pumping with detuning.
Superimposed in black is a line indicating where $\partial\omega/\partial t=0$ according
to \protect\refeq{de}.  Superimposed on this line are the solutions to this equation
which result as $\tau$ is scanned forward (yellow or light grey) and backward (white).}
\label{Cfig}
\end{figure}

Processes that change the total nuclear magnetization at the
high magnetic fields used here (4 to 10~T) are unlikely to be
due to the flip-flop terms of the contact hyperfine interaction
of the ground-state electron in the \QD, as its energy levels
are known to be narrow (on the order of $\hbar/T_2$, with
$T_2\sim 3~\mu$s) leaving few viable pathways for
energy-conserving nuclear-spin flips. In contrast, the dipolar
interaction between a trion's unpaired hole and a nuclear spin
may induce a spin-flip with the nuclear Zeeman energy
compensated by the broad width of the emitted photon
($\gamma/2\pi\sim 0.1$~GHz). Fermi's golden rule allows an
estimate of the rate at which a trion hole (at position
$\ts{\vec{r}}{h}$, with gyromagnetic ratio $\ts{g}{h}$)
polarized along the sample growth axis (orthogonal to the
magnetic field) randomly flips a nuclear spin at position
$\vec{r}$ in a spatially flat \QD\ during spontaneous emission,
with the photon energy density of states negligibly changed by
the Zeeman energy of the nucleus.  The result is
$\Gamma(\vec{r}) \approx
(9\mu_0^2/128\pi)(\tsc\mu{b}\ts{g}{h}/B_0)^2\gamma \langle
|\vec{r}-\ts{\vec{r}}{h}|^{-3}\rangle^2\sim 1/(20\text{~ms})$,
where the brackets refer to an average over the hole
wavefunction. Nuclear polarization due to this process has been
considered before in the modeling of similar effects
\cite{Reinecke_nuclear,sham_cpt}. The change in the nuclear
magnetization induced by the trion is random; spins may flip in
either direction, leading to a random walk. However, the
\emph{rate} of this walk is itself modified by the probability
that the trion is excited.  This probability is a function of
the pulse sequence; for the sequence shown in \reffig{data}(b),
we take it as
\be
\label{countrate}
C(\omega,\tau)=\ts{S}{p}^z
\frac{[1-\exp(-\beta(\omega)T)][1-\cos[(\omega_0+\omega)\tau]]}{1-\cos[(\omega_0+\omega)\tau]\exp(-\beta(\omega)T)}.
\ee
Here, $\omega_0$ is the Larmor frequency of the electron spin
in the absence of nuclear shifts, $\omega$ is the randomly
drifting Overhauser shift, $\beta(\omega)$ is the rate of
optical pumping, $T$ is the pumping time, and $\ts{S}{p}$ is
the saturation value of the polarized spin, equal to $1/2$ for
perfect pumping.  This function is plotted in \reffig{Cfig}.
%The
%Overhauser shift $\omega$ depends on the average nuclear
%magnetization as $\omega=\int d^3\vec{r} A(\vec{r}) \am$, where
%$A(\vec{r})$ is the hyperfine field of the electron.
$C(\omega,\tau)$ oscillates sinusoidally with increasing $\tau$
due to the spin's Larmor precession; the Overhauser shift
$\omega$ affects the frequency of Larmor precession.  For large
values of $|\omega|$, the trion transition shifts away from
resonance with the optical pumping laser, leading to reduced
pumping efficiency and trion creation.
%The shape of
%$C(\omega,\tau)$ drives the feedback loop, in which
%The magnetization drift caused by trions changes $\omega$, in
%turn changing the rate of magnetization drift.
If magnetization drift due to trions were the only process,
$\omega$ would randomly drift to whatever value is needed to
null $C(\omega,\tau)$, thereby slowing and stopping the drift.
As a result, no fringes would be observed.

The second process which counters this drift is the presence of
nuclear spin diffusion.  When trion emission pushes $\omega$ to
too large a value, nuclear dipolar interactions ``flatten" the
nuclear magnetization.  As a result, the shift $\omega$ is
``pulled" back to a low value, countering the tendency of trion
emission to push $\omega$ away from zero. The stable
quasi-equilibrium value of $\omega$ resulting from the balance
of these processes lives on the edge of the fringes shown in
\reffig{Cfig}; the nuclear polarization ``surfs" along the edge
of this function as $\tau$ is changed.  As $\tau$ is increased,
$|\omega|$ increases causing the observable photon count to
decrease due to the reduced degree of optical pumping.  When
$|\omega|$ is so high that pumping is ineffective
($\beta(\omega)\rightarrow 0$) and the trion-induced walk
stops, spin-diffusion causes the system to drift back to a new
stable magnetization at a lower value of $|\omega|$, and the
process continues.

These processes may be formally modeled by a diffusion equation
for the nuclear distribution.  In this model, the nuclear
magnetization at each nuclear site $j$ is a random variable,
$M_j$. A probability distribution function (\pdf)
$f(m_1,m_2,\ldots;t)=f(\vec{m};t)$ gives the joint probability
that the nuclear magnetization at each position is $M_j=m_j$ at
time $t$.  The Overhauser shift is then also a random variable
$\Omega$, defined by
%The time-dependent average value of
%a general function $g[m(\vec{r})]$ of the nuclear magnetization
%is formally written $\langle g[m(\vec{r})](t)\rangle = \int
%\mathcal{D}[m(\vec{r})] g[m(\vec{r})] f[m(\vec{r}),t]$, where
%$\mathcal{D}[m(\vec{r})]$ provides a product of differentials
%for every location $\vec{r}$.
%The time-derivative of this
%\pdf\ has several terms.
$\Omega=\sum_j A(\vec{r}_j) M_j,$ where $A(\vec{r}_j)$ is the
electron hyperfine field at the position $\vec{r}_j$ of nucleus
$j$. The average value of $\Omega$ at time $t$ is written
$\langle \Omega \rangle = \omega(t)$ and is found by averaging
over all possible values of each $M_j$, weighted by the joint
\pdf\ $f(\vec{m};t)$. This joint \pdf\ obeys the equation
\begin{multline}
\label{pde}
\frac{\partial f}{\partial t} = \sum_j \biggl\{-\sum_{k} D_{jk}
\biggl(\frac{\partial f}{\partial m_k}-\frac{\partial f}{\partial m_j}\biggr) +\\
\biggl[F_j +\Gamma(\vec{r}_j) C(\Omega,t)\biggr]
\frac{\partial^2 f}{\partial m_j^2}\biggr\}.
\end{multline}
The first term, in which the sum over $k$ is the sum over
neighbors of $j$, describes the dissipative component of
nuclear spin diffusion with diffusion rates $D_{jk}$. The
second term describes the random walk of the magnetization at
each location $\vec{r}$ due to stochastic nuclear spin-flips
from the trion hole-spin; the constant $F_j$ models the
fluctuating component of nuclear spin diffusion, a term needed
to understand single \QD\ $T_2^*$ effects in the absence of the
nonlinearities we consider here.

A detail omitted from this model is the degree to which the
nuclear spin diffusion is slowed or directed by the presence of
the electronic hyperfine field.  The magnetic field gradient
introduced by the electron does not completely freeze diffusion
in this case, because the electronic wave-function is large in
comparison to the nuclear-nuclear spacing. Reference
\onlinecite{gr73} indicates methods by which this could be
modeled.  However, our work solving \refeq{pde} including such
methods added greater complication and little insight into the
observed data.

Remarkably, the data of \reffig{data} can be understood by
examining just the average shift $\omega(t)$ using two
simplifying assumptions. In the first assumption, the electron
wavefunction is modeled as a surface with a hard boundary, and
the first term of \refeq{pde} acts to decrease $\omega(t)$ by
modeling diffusion through that boundary.  This results in the
equation
\be
\label{halfway}
\frac{\partial\omega}{\partial t} = -\kappa \omega +
\alpha\left\langle \frac{\partial^2}{\partial \Omega^2} [\Omega
C(\Omega,t)]\right\rangle.
\ee
The constant $\kappa$ depends on the electronic wavefunction
and the rate of nuclear diffusion, but we treat this parameter
as adjustable rather than attempting a microscopic description.
The constant $\alpha$ is formally given by $\sum_j
\Gamma(\vec{r}_j) A^2(\vec{r}_j)$.

Unfortunately, \refeq{halfway} is not a closed system of
equations, because it still requires full knowledge of
$f(\vec{m};t)$ to solve.  However, if $C(\Omega,\tau)$ is a
sufficiently flat function of $\Omega$ in comparison to the
width of $f(\vec{m};t)$, then we may treat $C(\Omega,\tau)$ as
roughly constant at $C(\omega(t),\tau)$ over the small width of
$f(\vec{m};t)$.  This is our second assumption, leaving the
final equation
\be
\frac{\partial\omega}{\partial t} = -\kappa \omega +
\alpha\frac{\partial^2}{\partial \omega^2} [\omega
C(\omega,t)].
\label{de}
\ee
Invoking our separation of timescales, we presume $\omega(t)$
evolves from its initial value (set by the last chosen value of
$\tau$) to a quasi-equilibrium final value $\ts\omega{f}$. This
final value determines the expected count rate
$C(\ts{\omega}{f},\tau)$ at this value of $\tau$.  We solve by
assuming $\omega(0)=0$ at the first attempted value of $\tau$,
and then we scan $\tau$ up and then down as in the experiment,
finding the steady-state solution of \refeq{de} at each value.

\begin{figure}
\includegraphics[width=0.8\columnwidth]{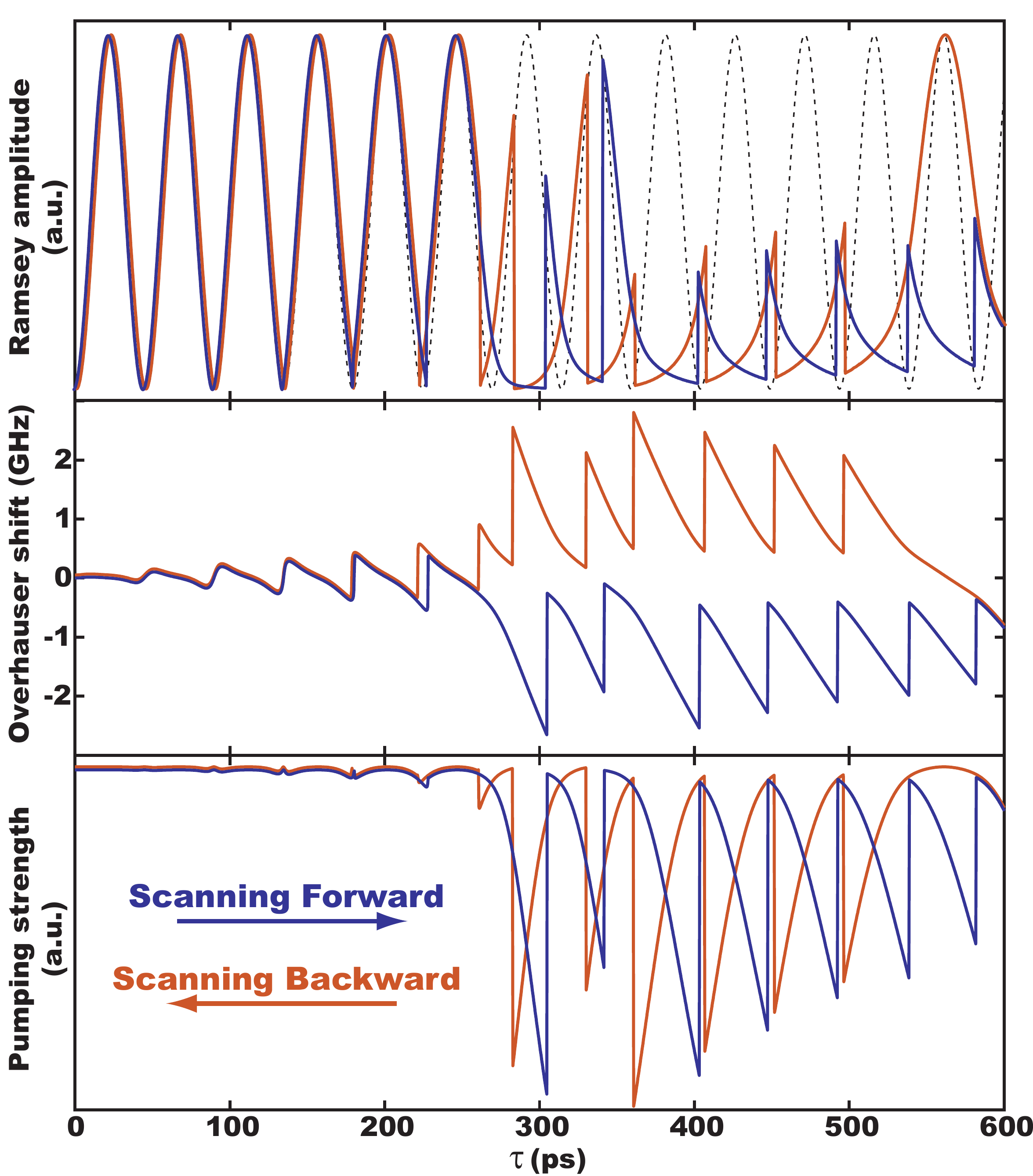}
\caption{(Color online) The modeled (a) countrate or Ramsey amplitude $C(\ts\omega{f},\tau)$,
(b) Overhauser shift $\ts\omega{f}$, and (c) Pumping strength $\beta(\ts\omega{f})$.
The dotted line in (a) is the expected Ramsey fringe in
the absence of nuclear effects.
The traces in (b) are the same as those in Fig.~\ref{Cfig}.
The blue or darker grey line corresponds to scanning $\tau$ forward, and the red or
lighter grey line corresponds to scanning $\tau$ backward. }
\label{model_traces}
\end{figure}

Figure~\ref{model_traces} shows the modeled
$C(\ts{\omega}{f},\tau)$, $\ts\omega{f}$, and
$\beta(\ts\omega{f})$ as a function of $\tau$.  This particular
model used $\kappa/\alpha=10^4$, which reproduces the
qualitative shape of the data quite well, and quantitatively
reproduces the location where sinusoidal fringes evolve into
sawtooth-like fringes.
%Detailed curve fitting is
%not reasonable for this process, since the quantitative
%behavior depends on a random initial condition.
Qualitative differences are dominated by the random conditions
that develop when the stage is moved on its rail, forming the
breaks between data sets in \reffig{data}(a).  Details of the
shape of the waveform are related to the assumed form of the
optical absorption.  For simplicity, we have used
$\beta(\omega)=\beta_0\exp(-\omega^2/2\sigma^2)$, with
$\sigma/2\pi=1.6$~GHz and $\beta_0=3/T$ for known pumping time
$T=26$~ns, which roughly matches the experimentally observed
count-rate when scanning the pump laser across the resonance.
The real absorption shape is difficult to observe directly
since hysteretic nuclear pumping effects also appear in
absorption experiments with scanning CW lasers, as reported
elsewhere~\cite{sham_cpt,imamoglu_cpt}.

This effect may be useful for future coherent technologies
employing \QDs.   This pulse sequence may serve as a
``preparation step" for a qubit to be used in a quantum
information processor, as it tunes the qubit to a master
oscillator~\cite{fast_rotations_prop} and narrows the random
nuclear distribution, assisting more complex coherent
control~\cite{vandersypen_singlet-triplet,Reilly_Zamboni_Science,delftlock,new_yacoby,greilich_nuclear,Reinecke_nuclear}.
In particular, the ability to control a single electron with
effectively $\delta$-function-like rotation pulses introduces
strong potential for dynamical decoupling~\cite{udd,lidar08},
but many schemes, especially those that compensate for pulse
errors such as the Carr-Purcell-Meiboom-Gill (CPMG) sequence,
require some method to tune the \QD's Larmor period to an
appropriate division of the pulse-separation time.

In conclusion, we have observed nonlinear nuclear feedback
effects in a single charged \QD\ resulting from the countering
processes of random nuclear walks driven by trion creation, the
finite width of optical absorption, Overhauser-shifted Larmor
precession, and nuclear spin diffusion.  This feedback may be
employed for tuning the electron Larmor period to a particular
pulse separation time for more complex pulses sequences.
Although the model we have presented is simple and replicates
the data well, it is insufficient to describe the processes in
a \QD\ under all possible pulse sequences. In particular, these
nonlinear effects are highly suppressed in spin-echo
measurements~\cite{pressecho}, even though trion creation
follows a similar nonlinear function to \refeq{countrate}.
Future work will involve extending this model to explain the
non-Markovian effects of nuclei under more complex pulse
sequences, as well as exploiting it to extend \QD-based quantum
memories.

We thank Erwin Hahn for valuable discussions. This work was
supported by \textsc{NICT, MEXT, NSF CCF0829694}, and Special
Coordination Funds for Promoting Science and Technology. PLM
was supported by the David Cheriton Stanford Graduate
Fellowship.

%\bibliography{../bib/bibstrings,%
%              ../bib/quantumdots,%
%              ../bib/NMR}

\end{document}